\documentclass[aps,prl,preprint,showpacs,superscriptaddress]{revtex4-1}

\usepackage{graphicx}
\usepackage{newtxtext,amssymb}

\usepackage{color}

\begin{document}

\title{
	Shrinking of Rapidly Evaporating Water Microdroplets Reveals their Extreme Supercooling
	}

\author{Claudia Goy}
\affiliation{Institut f\"ur Kernphysik, J. W. Goethe-Universit\"at Frankfurt(M), 60438 Frankfurt(M), Germany}
\author{Marco A. C. Potenza}
\affiliation{Dipartimento di Fisica, Universit\`a degli Studi di Milano, 20133 Milano, Italy}
\author{Sebastian Dedera}
\affiliation{Institute of Earth Sciences, 69120 Heidelberg, Germany}
\author{Marilena Tomut}
\affiliation{GSI - Helmholtzzentrum f\"ur Schwerionenforschung, 64291 Darmstadt, Germany}
\author{Emmanuel Guillerm}
\affiliation{Univ Lyon, Universit\'e Claude Bernard Lyon 1, CNRS, Institut Lumi\`ere Mati\`ere, 69622 Lyon, France}
\author{Anton Kalinin}
\affiliation{Institut f\"ur Kernphysik, J. W. Goethe-Universit\"at Frankfurt(M), 60438 Frankfurt(M), Germany}
\affiliation{GSI - Helmholtzzentrum f\"ur Schwerionenforschung, 64291 Darmstadt, Germany}
\author{Kay-Obbe Voss}
\affiliation{GSI - Helmholtzzentrum f\"ur Schwerionenforschung, 64291 Darmstadt, Germany}
\author{Alexander Schottelius}
\affiliation{Institut f\"ur Kernphysik, J. W. Goethe-Universit\"at Frankfurt(M), 60438 Frankfurt(M), Germany}
\author{Nikolaos Petridis}
\affiliation{GSI - Helmholtzzentrum f\"ur Schwerionenforschung, 64291 Darmstadt, Germany}
\author{Alexey Prosvetov}
\affiliation{GSI - Helmholtzzentrum f\"ur Schwerionenforschung, 64291 Darmstadt, Germany}
\author{Guzm\'an Tejeda}
\affiliation{Laboratory of Molecular Fluid Dynamics, Instituto de Estructura de la Materia, CSIC, 28006, Madrid, Spain}
\author{Jos\'e M. Fern\'andez}
\affiliation{Laboratory of Molecular Fluid Dynamics, Instituto de Estructura de la Materia, CSIC, 28006, Madrid, Spain}
\author{Christina Trautmann}
\affiliation{GSI - Helmholtzzentrum f\"ur Schwerionenforschung, 64291 Darmstadt, Germany}
\affiliation{Material- und Geowissenschaften, Technische Universit\"at Darmstadt, 64287 Darmstadt, Germany}
\author{Fr\'ed\'eric Caupin}
\affiliation{Univ Lyon, Universit\'e Claude Bernard Lyon 1, CNRS, Institut Lumi\`ere Mati\`ere, 69622 Lyon, France}
\author{Ulrich Glasmacher}
\affiliation{Institute of Earth Sciences, 69120 Heidelberg, Germany}
\author{Robert E. Grisenti}\email[]{grisenti@atom.uni-frankfurt.de}
\affiliation{Institut f\"ur Kernphysik, J. W. Goethe-Universit\"at Frankfurt(M), 60438 Frankfurt(M), Germany}
\affiliation{GSI - Helmholtzzentrum f\"ur Schwerionenforschung, 64291 Darmstadt, Germany}

%\date{\today}

\begin{abstract}

The fast evaporative cooling of micrometer-sized water droplets in vacuum offers the appealing possibility to investigate supercooled water -- below the melting point but still a liquid -- at temperatures far beyond the state-of-the-art. However, it is challenging to obtain a reliable value of the droplet temperature under such extreme experimental conditions. Here, the observation of morphology-dependent resonances in the Raman scattering from a stream of perfectly uniform water droplets has allowed us to measure with an absolute precision of better than 0.2\% the variation in droplet size resulting from evaporative mass losses. This finding proved crucial to an unambiguous determination of the droplet temperature. In particular, a fraction of water droplets with initial diameter of $6379\pm 12$~nm were found to remain liquid down to $230.6\pm 0.6$~K. Our results question temperature estimates reported recently for larger supercooled water droplets, and provide valuable information on the hydrogen-bond network in liquid water in the hard-to-access deeply supercooled regime.

\end{abstract}

\maketitle

Water can exist in the liquid state at temperatures far below its normal melting point. The first report on the observation of supercooled water probably dates back to Fahrenheit, who had cooled water to 264~K \cite{Fahrenheit1724}. A better understanding of the properties of supercooled water \cite{Poole1992,Palmer2014}, as well as establishing how and at which rates it transforms to ice \cite{Moore2011,Russo2014,Haji-Akbari} represent important goals with potentially broader impacts. For instance, tiny droplets of supercooled water at temperatures as low as 238~K naturally occur in the upper clouds of Earth's atmosphere \cite{Rosenfeld2000}, and an improved description of atmospheric ice formation could help to develop more reliable climate models \cite{Murray2012}. More generally, the structural properties of supercooled water have been related to its anomalous behavior \cite{Nilsson2015}. Water is an unusual liquid because many dynamic properties (such as the viscosity and relaxation times) and thermodynamic response functions (such as the heat capacity and the isothermal compressibility) show a power-law increase that becomes more pronounced in the supercooled state \cite{Debenedetti2003}, suggesting an apparent singularity at an estimated temperature of $\approx 228$~K \cite{Speedy76,Dehaoui2015}. Yet what sort of singularity might water be approaching still represents an unresolved puzzle that has prompted the formulation of conflicting scenarios to interpret its origin \cite{Gallo2016}. 

The deeply supercooled regime is experimentally difficult to investigate due to the rapidly increasing ice nucleation rate with decreasing temperature. Conventional techniques such as those based on the use of thin capillaries \cite{Hare1986} or emulsions \cite{Taborek1985,Riechers2013} allow studying supercooled water only at temperatures above $\approx 235$~K. In an effort to access lower temperatures, ice formation has been prevented in experiments with nano-confined water \cite{Mallamace2013}, aqueous solutions \cite{Murata2012} and nanometer-scale water clusters \cite{Manka2012}. However, the question of how the results of all these studies extrapolate to supercooled bulk water remains controversial \cite{Caupin2015}.

Micrometer-sized water droplets formed in a laminar liquid jet in vacuum offer a promising strategy to investigate supercooled bulk water and ice formation at very low temperatures \cite{Faubel1988,Sellberg2014,Sellberg2015,Laksmono2015}. This is due to the combination of small sample size, suppression of heterogeneous nucleation sites, and fast evaporative cooling, with cooling rates exceeding $\sim 10^6$~K s$^{-1}$ for sub-10 $\mu$m-diameter water droplets. Despite its obvious benefits, this approach still lacks sufficient reliability in the determination of the droplet temperature. One conventional way of estimating the temperature of a liquid water jet in vacuum is based on the Knudsen kinetic model of evaporative cooling \cite{Faubel1988,Smith2006,Sellberg2014}. This model, however, depends critically on key experimental parameters, such as the droplet diameter, which are difficult to determine with sufficient precision. Here, by Raman spectroscopy of a microscopic water jet we have obtained a precise measure of the evolution in the size of the rapidly evaporating droplets, thus establishing accurately their temperature and providing unambiguous evidence for the existence of supercooled bulk water down to 230.6~K.

\begin{figure}[t]
\includegraphics[width=1\linewidth]{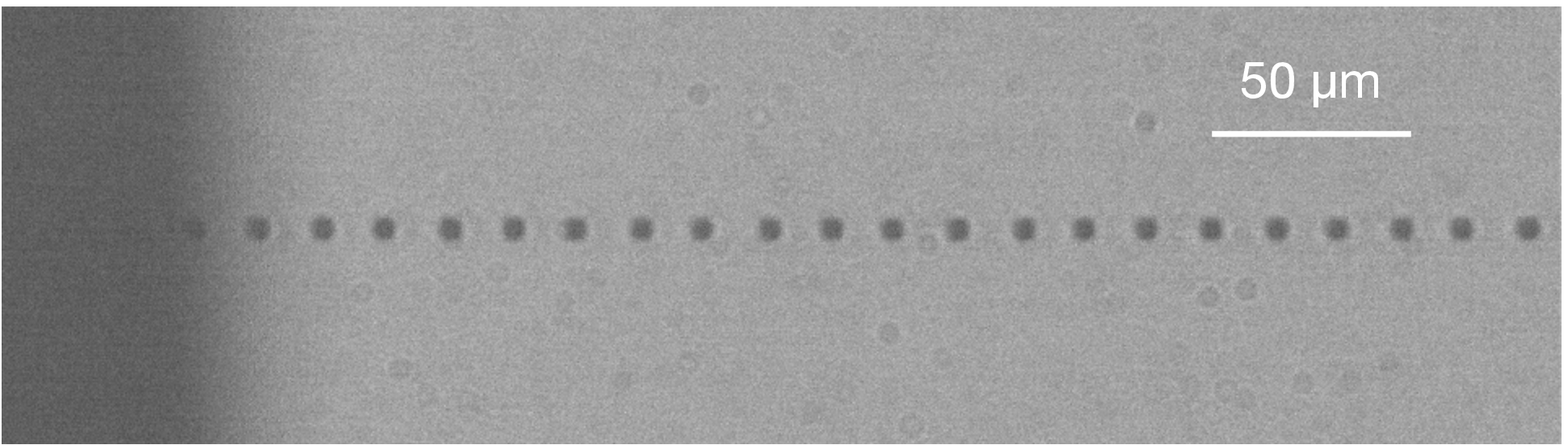}
\caption{\label{}Stroboscopic image of the liquid water jet emerging from a $3.16\pm 0.1$~$\mu$m-diameter glass capillary nozzle, whose shadow is visible on the left. The jet breakup was triggered by an external excitation at the frequency of $966$~kHz to produce a periodic stream of perfectly uniform water droplets.}
\label{Fig_Jet}
\end{figure}

For a water droplet whose volume $V(t)$ and temperature $T(t)$ change with time $t$ the probability $\Sigma(T)$ to remain liquid at the temperature $T$ below the melting point $T_m$ is given by \cite{Riechers2013,Azouzi2013}
\begin{equation}
\ln\Sigma(T)\simeq -V(0)\int_{T}^{T_m}\frac{J\left(T^{\prime}\right)}{\vert\dot{T^{\prime}}\vert}dT^{\prime},
\label{Prob}
\end{equation}
where $J(T)$ is the rate of ice nucleation and $\vert\dot{T}\vert$ is the cooling rate. Accordingly, the smaller the droplet size, the higher the degree of supercooling that can be reached. While liquid water jets emerging from orifices with diameter $d\gtrsim 10$~$\mu$m are routinely produced in the laboratory, nozzle clogging had severely limited the use of smaller apertures~\cite{DePonte2008}. The water jets investigated here were produced by injecting ultra-pure liquid water at $292.3\pm 0.1$~K into a vacuum chamber through a $d=3.2\pm 0.1$~$\mu$m-diameter glass capillary nozzle. We have generated a periodic stream of perfectly uniform droplets (Fig.~\ref{Fig_Jet}) by applying an external excitation at the frequency $f=966$~kHz by means of a piezoelectric actuator to trigger the jet breakup \cite{Frohn2000}.

The liquid water jet in Fig.~\ref{Fig_Jet} was probed by recording Raman spectra of the O$-$H stretching mode as function of the distance $z$ from the nozzle. The experimental setup was similar to that described in Ref.~\cite{Kuehnel2011}. In the present work, we have employed a portable Raman instrument (iHR320, Horiba Jobin Yvon) consisting of a head for in-situ analysis fiber-coupled to a grating spectrometer equipped with a high sensitivity CCD detector, and providing a resolution of $\approx 1.5$~cm$^{-1}$. As excitation source we used an external 6 W Nd:YVO3 cw-laser (Verdi V6, Coherent), which we operated up to a maximum power of 2.5~W, generating a radiation beam at $\lambda_0=532$~nm that was focused to a $\approx 25$~$\mu$m-diameter spot onto the liquid water jet in the vacuum chamber. The Raman signal was recorded under an angle of 90$^{\circ}$ with respect to the excitation laser beam and focused to the entrance of the Raman instrument's head. The Raman shifts $\Delta \nu=1/\lambda_0-1/\lambda$, where $\lambda$ is the scattered wavelength, were calibrated by using a silicon sample. The entire liquid jet source was mounted on micro-actuators stages that allowed a displacement of the jet along $z$ with a precision of $\approx 1$~$\mu$m.

\begin{figure}[t]
\includegraphics[width=1\linewidth]{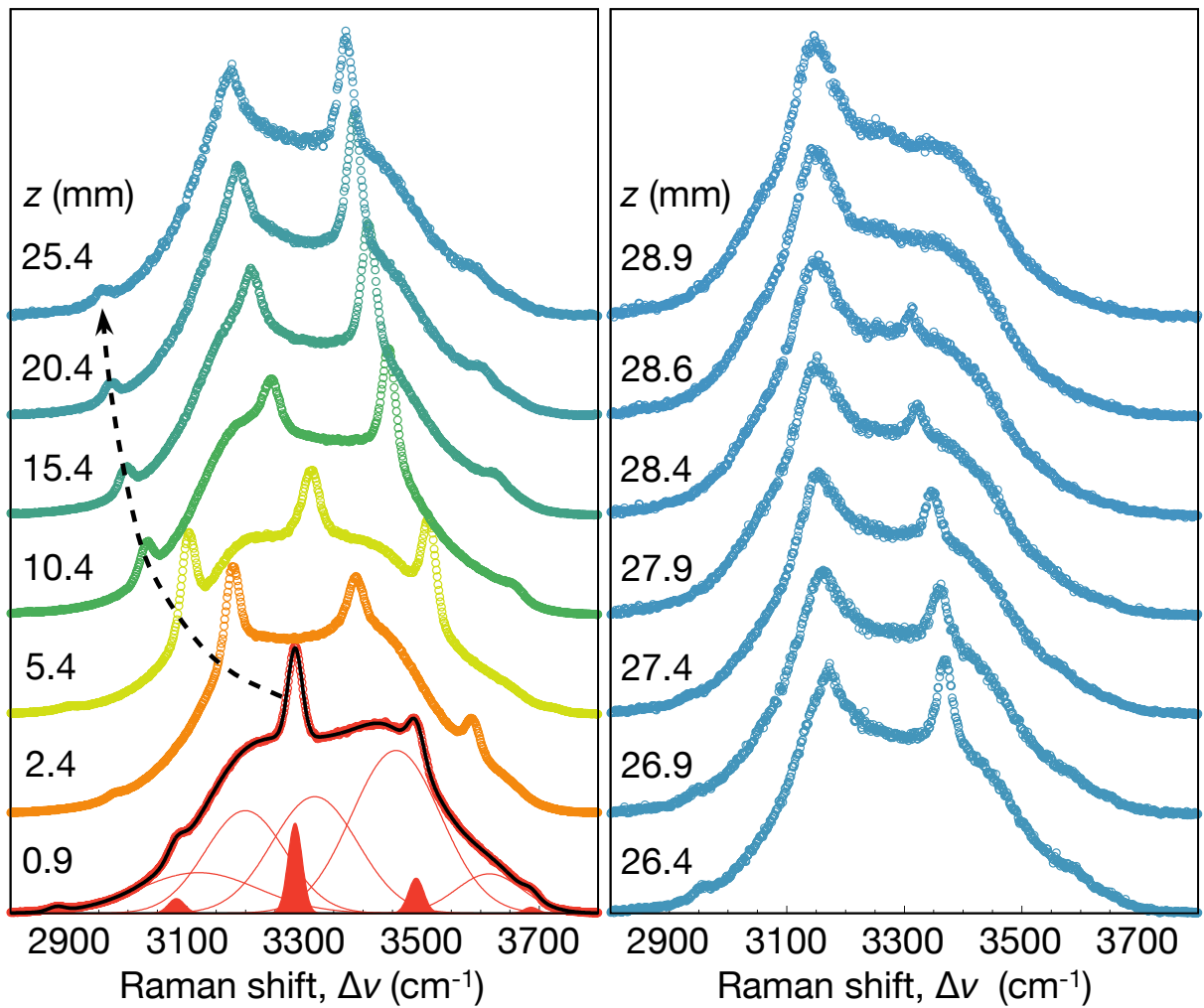}
\caption{\label{}Normalized and two-points baseline-corrected Raman spectra measured as function of the distance from the nozzle, indicated by labels in both panels. The thick black solid curve is a fit to the experimental spectrum at $z=0.9$~mm by assuming five Gaussian functions for the O$-$H stretching band (shown as thin solid lines) and five additional Gaussian peaks for the resonances (shown as filled curves).}
\label{Fig_Raman}
\end{figure}

Selected Raman spectra measured at distances between $z=0.9$ and 28.9~mm from the nozzle are shown in Fig.~\ref{Fig_Raman}. During a typical acquisition time of 300~s up to $\sim 10^9$ individual droplets crossed the exciting laser beam focus. The spectrum at 28.9 mm in the right panel of Fig.~\ref{Fig_Raman} coincides with that from bulk crystalline ice \cite{Sun2013}, indicating that within our resolution all droplets have frozen to ice at the largest distance investigated here.

The most striking feature visible in Fig.~\ref{Fig_Raman} is the presence of up to five narrow peaks superimposed on the O$-$H stretching bands, which progressively shift to smaller wave numbers with increasing distance from the nozzle. These peaks originate from a morphology-dependent resonant enhancement of the Raman scattering for specific values of the droplet radius-to-wavelength ratio \cite{Thurn1985,Kiefer1996}. The resonances can be viewed as standing waves at the droplet-vacuum interface and require for their occurrence a perfectly smooth spherical shape. This explains the absence of the resonance peaks in the Raman spectrum from frozen droplets, which will tend to exhibit a more irregular interface with respect to the smooth surface of a liquid droplet. Accordingly, the attenuation of the resonance peaks observed at the largest distances (right panel in Fig.~\ref{Fig_Raman}) can be interpreted as due to the rapidly decreasing fraction of purely liquid water droplets. We further note that by turning off the piezo actuator the resonance peaks invariably disappeared, thereby confirming the uniformity of droplet sizes for the piezo-driven jet breakup (Fig.~\ref{Fig_Jet}).

We now show that the observation of the resonances offers the most accurate and precise way to determine the droplet diameter \cite{Kiefer1996} and, in turn, the droplet temperature. The resonances are described in the framework of the Mie-Debye light scattering theory. Relevant to the present discussion is the ratio between the extinction and the geometrical ($\pi r^2$, with $r$ the droplet radius) cross sections given by \cite{Chylek76,Chylek78,Bohren98}
\begin{equation}
Q_{\rm ext}(x,n)=\frac{2}{x^2}\sum_{m=1}^{\infty}(2m+1)\mathrm{Re}\left[a_m(x,n)+b_m(x,n)\right],
\label{Qext}
\end{equation}
where $x = 2\pi r/\lambda$ is the size parameter, $n$ is the refractive index, and $a_m(x,n)$ and $b_m(x,n)$ are the complex partial-wave expansion scattering amplitudes. Within the size range of interest here the function $Q_{\rm ext}(x,n)$ exhibits a smooth ripple structure with an infinite series of sharp peaks occurring at definite values $\left\{x_i\right\}_{i\in \mathbb{N}}$ of the size parameter \cite{Chylek76, Chylek78}. Each value corresponds to a resonance condition for a specific radius-to-wavelength ratio. Accordingly, if $x_i$ is the size parameter associated with any of the observed resonances centered at $\Delta \nu_j$, $j=0,\ldots,4$, then we have
\begin{equation}
\Delta \nu_j = \frac{1}{\lambda_0} - \frac{x_i}{2\pi r}.
\label{shift}
\end{equation}
Equation~(\ref{shift}) relates the observed shift to smaller wave numbers of the resonance peaks to a reduction of the droplet radius with increasing distance from the nozzle. There is no unambiguous way to assign the size parameter $x_i$. However, it can be shown that $\forall i\in \mathbb{N}$ the spacing $\Delta x=x_{i+2}-x_i$ is only a very slowly varying function of the refractive index \cite{Chylek76}. For two observed resonances centered at $\Delta \nu_j$ and $\Delta \nu_{j+2}$ we thus obtain from Eq.~(\ref{shift}) for the droplet diameter
\begin{equation}
D=\frac{1}{\pi}\frac{\Delta x}{\Delta \nu_{j+2}-\Delta \nu_j}.
\label{D}
\end{equation}
The Raman shifts $\Delta \nu_j$ were determined by fitting each Raman spectrum to five broad Gaussian components representing the fundamental O$-$H stretching band \cite{Suzuki2012,Sun_b2013} and up to five additional Gaussian peaks representing the resonances; one example of such a fit is shown in Fig.~\ref{Fig_Raman} as black solid curve for the spectrum measured at $z=0.9$ mm (left panel).

The direct numerical evaluation of $Q_{\rm ext}(x,n)$ [Eq.~(\ref{Qext})] shows that $\Delta x= 1.6396 - 0.6312n$ with standard deviation of $9.4\times 10^{-4}$ in the range from $n=1.333$, the refractive index of liquid water at the normal melting point and $\lambda=632$~nm \cite{Harvey1998}, down to $n=1.315$ as extrapolated for supercooled water at $\approx 230$~K. For the extrapolation we adopted a modified Clausius-Mossotti relation as in the formulation by the International Association for the Properties of Water and Steam (IAPWS), keeping the linear terms in the temperature and density $\rho$, i.~e., $(n^2-1)/[(n^2+2)\rho]=a_0+a_1T+a_2\rho$ \cite{Harvey1998}. The wavelength dependence of the refractive index across the spectral Raman O$-$H stretching band can be safely neglected here \cite{Harvey1998}. The coefficients $a_0=0.230278$, $a_1=-1.1137\times 10^{-5}$, and $a_2=-0.0245171$ were determined by fitting the IAPWS expression to experimental data for the refractive index down to 258~K \cite{Carroll2002}. For the temperature dependence of the density of water we have assumed the sixth-order polynomial reported in Ref.~\cite{Hare1987}.

\begin{figure}[t]
\includegraphics[width=1\linewidth]{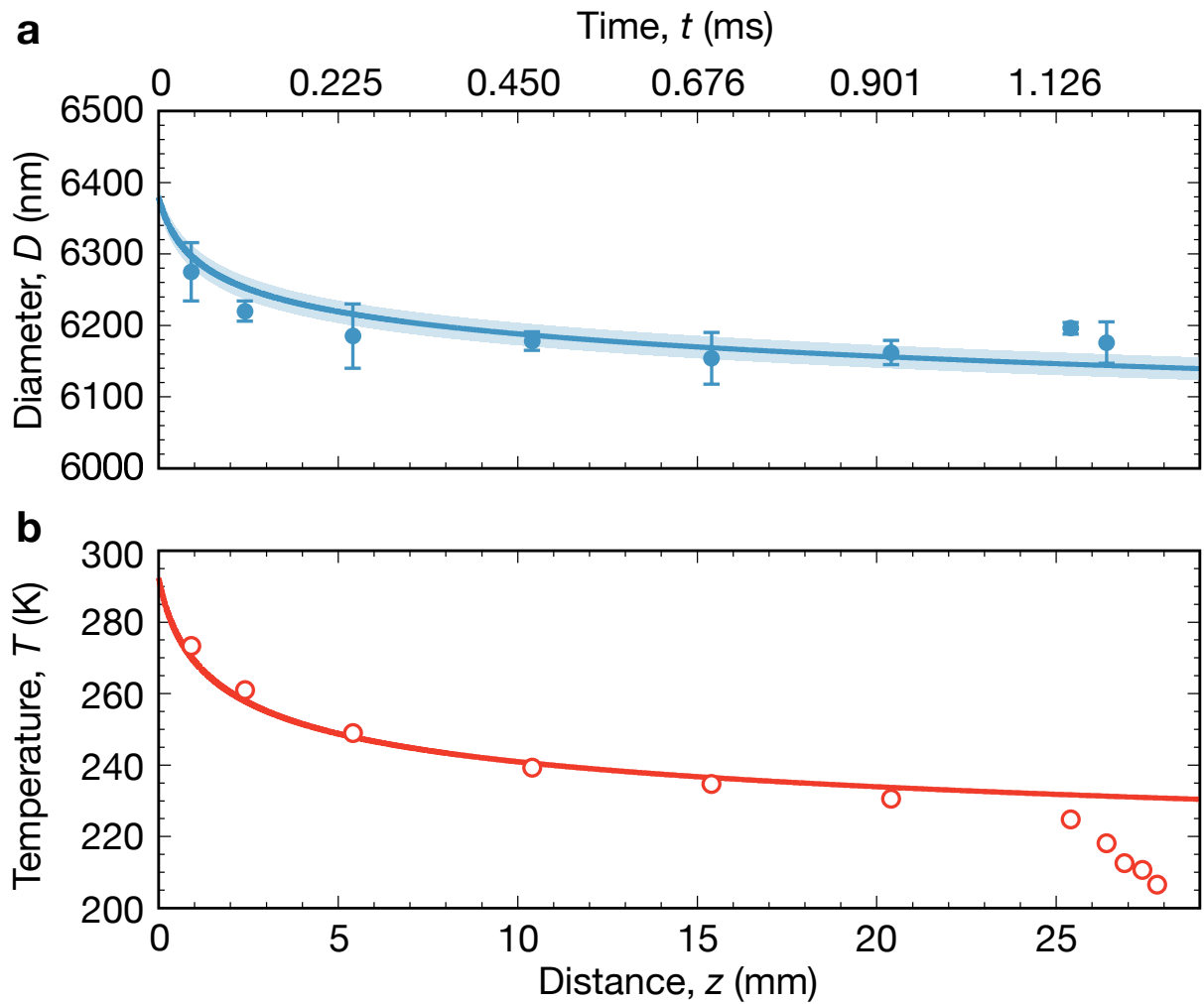}
\caption{\label{}Droplet diameter and temperature as function of the distance from the nozzle, i. e., travel time $t=z/v$, with $v=22.2$~m~s$^{-1}$ (upper $x$-axis). (a) The filled circles represent the droplet diameter determined by an iterative approach from the positions of the resonance peaks in the Raman spectra according to Eq.~(\ref{D}). The thick solid curve is a fit to the data based on the Knudsen model of evaporative cooling. The light-shaded region indicates the 68\% confidence interval. (b) The thick solid curve is the calculated volume-averaged droplet temperature corresponding to the fit shown in (a). The temperature uncertainty is represented by the line thickness. The open circles represent the droplet temperature obtained from the analysis of the shape of the Raman O$-$H stretching bands shown in Fig.~\ref{Fig_OH_bands}. The error bars are comparable to the symbol size.}
\label{Fig_D_and_T}
\end{figure}

In order to determine the droplet diameter as a function of the distance from the nozzle on the basis of Eq.~(\ref{D}) one has to know the droplet temperature, which is the unknown variable that we aimed at establishing. To circumvent this problem we adopted an iterative approach. Starting with an arbitrary constant value of the refractive index we evaluated the droplet diameter by averaging Eq.~(\ref{D}) over the pairs of resonance peaks visible in each Raman spectrum. We performed a chi-square fit to the obtained values by using as the fit function the droplet diameter $D(z)$ computed by the Knudsen model of evaporative cooling with the initial droplet diameter $D(0)\equiv D_0$ and the jet velocity $v$ as fit parameters. In the calculations we have taken into account the initial temperature gradient throughout the droplet resulting from the finite thermal conductivity of liquid water \cite{Smith2006,Sellberg2014}. The corresponding volume-averaged droplet temperature $T(z)$ was then used to establish a new set of values of the refractive index. The above steps were repeated until convergence was reached, and the resulting values for the droplet diameter are shown as filled circles in Fig.~\ref{Fig_D_and_T}(a). The thick solid line in Fig.~\ref{Fig_D_and_T}(a) is the model fit with $D_0=6379\pm 12$~nm and $v=22.2\pm 1.5$~m~s$^{-1}$, with the 68\% confidence interval indicated as light-shaded region. The thermodynamic functions of the Knudsen model were extrapolated from available experimental data at higher temperatures. In particular, for the latent heat of vaporization and the vapor pressure were assumed the expressions reported in Ref.~\cite{Murphy2005}, the thermal conductivity and surface tension were determined as discussed in Ref.~\cite{Sellberg2014}, and the isobaric heat capacity was derived from the vapor pressure by means of the Clausius-Clapeyron equation as shown in Ref.~\cite{Kalova2010}. We have verified that the result of the fit in the present temperature range was largely insensitive to the particular empirical extrapolations as discussed in Ref.~\cite{Sellberg2014}.

The quality of the fit in Fig.~\ref{Fig_D_and_T}(a) validates the Knudsen model and the choice of the extrapolated thermodynamic parameters, and thus provides a very accurate description of the droplet evaporative cooling. We note that the values of the fit parameters are consistent with (albeit much more precise than) $D_0=6560\pm 370$~nm and $v=18.3\pm 2.0$~m~s$^{-1}$ as inferred directly from the stroboscopic image in Fig.~\ref{Fig_Jet} by using standard equations relating $D_0$ to $d$, $f$, and $v$ \cite{Frohn2000}. The droplet temperature corresponding to the fit in Fig.~\ref{Fig_D_and_T}(a) is shown as thick solid curve in Fig.~\ref{Fig_D_and_T}(b), with the line thickness representing the range of uncertainty. At the largest distance of 28.4~mm at which resonances are still visible in the Raman spectra (Fig.~\ref{Fig_Raman}), i.e., a non negligible fraction of droplets are still liquid, we infer a droplet temperature of $230.6\pm 0.6$~K. This value represents the lowest temperature established unambiguously for supercooled bulk water.

\begin{figure}[t]
\includegraphics[width=1\linewidth]{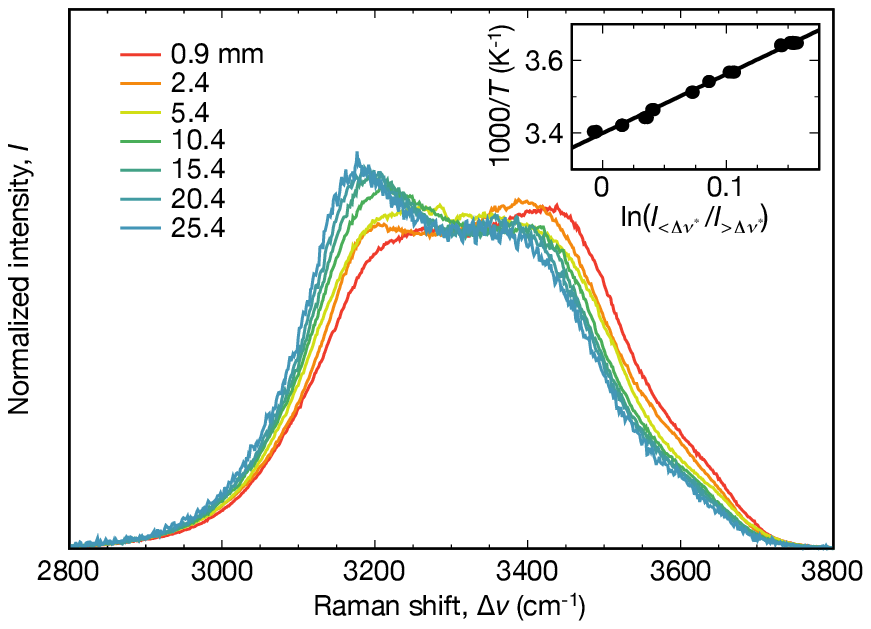}
\caption{\label{}Fundamental O$-$H stretching bands obtained by subtracting the contribution of the resonances from the Raman spectra in Fig.~\ref{Fig_Raman}. For clarity, only the bands up to $z=25.4$~mm are shown. The inset shows the calibration data (filled circles) plotted as the inverse of the temperature against the natural logarithm of the ratio of the integrated band intensities $I_{<\Delta \nu^{*}}$ and $I_{>\Delta \nu^{*}}$ below and above $\Delta \nu^{*}=3360$~cm$^{-1}$, respectively. The solid line is a linear fit that was extrapolated to the supercooled region in order to determine the droplet temperature from the bands in the main figure.}
\label{Fig_OH_bands}
\end{figure}

For the purpose of comparison we have determined the droplet temperature also by the more conventional approach based on the analysis of the Raman O$-$H stretching band profiles. Since the O$-$H stretch vibration is a probe of the local hydrogen-bond network, it exhibits a strong variation with temperature \cite{Walrafen1967}. This dependence has been exploited in the past to estimate the temperature of supercooled water droplets investigated by Raman spectroscopy \cite{Smith2006}. Figure ~\ref{Fig_OH_bands} displays the O$-$H stretching bands obtained from the Raman spectra in Fig.~\ref{Fig_Raman} by subtracting the resonance peaks contribution. To determine the droplet temperature we have produced a calibration curve, shown in the inset in Fig.~\ref{Fig_OH_bands}, by recording Raman O$-$H stretching bands of liquid water enclosed in a 1 cm-wide glass cell connected to a thermostat in the temperature range 274 to 294~K. The inverse of the temperature plotted versus the natural logarithm of the ratio of the integrated band intensities below and above an arbitrary point close to the centre of the O$-$H stretching band yields a linear relationship that we extrapolated to the supercooled liquid state \cite{Smith2006}. Additional Raman measurements performed on a liquid water sample contained in a $\approx 20$ $\mu$m-inner diameter glass capillary tube by using a second Raman instrument at the Institut Lumi\`ere Mati\`ere have confirmed the validity of the established linear relationship down to 238.4~K.

The droplet temperature estimated from the variation of the shape of the Raman O$-$H stretching bands is shown in Fig.~\ref{Fig_D_and_T}(b) as open circles. We find a good agreement with the temperature curve obtained from the analysis of the resonance peaks up to $z=20.4$~mm. The deviations observed for $z\gtrsim 25$~mm are likely due to the growing contribution of the scattering from droplets that have frozen to ice at such large distances, affecting the shape of the O$-$H stretching band. We note that the agreement in Fig.~\ref{Fig_D_and_T}(b) implicitly extends the range of validity of the Raman-temperature calibration curve (Fig.~\ref{Fig_OH_bands}) down to $\approx 234$~K. Overall, the established consistency between the two distinct approaches provides a rigorous proof of the reliability of our droplet temperature determination.

The lowest droplet temperature reported here for $\approx 6$ $\mu$m-diameter droplets [Fig.~\ref{Fig_D_and_T}(b)] is not consistent with recent temperature estimates based on the Knudsen model for nominal $\approx 12$ $\mu$m-diameter water droplets probed with ultrashort X-ray laser pulses \cite{Sellberg2014}. Indicating with $\Sigma_{12\mu{\rm m}}$ and $\Sigma_{6\mu{\rm m}}$ the probabilities to observe 12 and 6~$\mu$m-diameter water droplets, respectively, in the supercooled state at 230.6~K, and by taking into account that the evaporative cooling rate increases with decreasing droplet diameter, from Eq.~(\ref{Prob}) it follows that $\Sigma_{12\mu{\rm m}}\lesssim \left(\Sigma_{6\mu{\rm m}}\right)^8\ll \Sigma_{6\mu{\rm m}}\ll 1$. However, in Ref.~\cite{Sellberg2014} it was estimated that nearly 100\% of the 12~$\mu$m-diameter droplets were liquid at 230.6~K, and a fraction of them were even reported to have supercooled further to 227~K. We note that effects of laser-induced droplet heating in our experiments were completely negligible because of the extremely short droplet transit time across the excitation laser beam of $\approx 10^{-6}$~s and the small absorption cross section at 532~nm of 0.0447~m$^{-1}$. Although ice nucleation can be triggered by short, 532~nm laser pulses at a high intensity threshold of $\sim 10^{16}$~W~m$^{-2}$ \cite{Aliotta2014}, this phenomenon is unlikely to have occured in the present study due to the several orders of magnitude lower laser intensity of $\approx 5\times 10^9$~W~m$^{-2}$. Thus, while the above apparent discrepancy can be definitively resolved only by a direct comparison between the two (Raman and X-ray) scattering techniques, our results indicate that the degree of supercooling of micrometer-sized water droplets investigated recently might be largely overestimated, thereby challenging the interpretation of the reported experimental data \cite{Sellberg2014,Sellberg2015,Laksmono2015,Pathak2016}. 

Vibrational spectroscopy has been widely applied to the study of the structure of bulk liquid water \cite{Bakker2010}. In particular, spectroscopic investigations of liquid water from ambient to supercooled conditions had evidenced a continuous evolution in the Raman spectral features \cite{Hare1990,Suzuki2012,Sun_b2013}. With decreasing temperature the low-frequency side of the O$-$H stretching band around 3200~cm$^{-1}$ becomes more pronounced with respect to the high-frequency side around 3400~cm$^{-1}$. One interpretation attributes this behavior to the change upon cooling in the population of two distinct local hydrogen-bond structures -- distorted and tetrahedral -- associated with the high- and low-frequency spectral branches, respectively \cite{Nilsson2015}. The Raman O$-$H stretching bands shown in Fig.~\ref{Fig_OH_bands} clearly confirm this trend, indicating that it further extends down to at least $\approx 232$~K. No definite conclusion can be drawn here on whether this trend continues at even lower temperatures due to the scattering from frozen droplets for $z\gtrsim 25$~mm. More insights in this respect may come from the low-energy vibrations involving intermolecular hydrogen bonds. Recent time-resolved optical Kerr effect measurements have identified clear signatures of two structural components in the low-frequency spectral region around 200~cm$^{-1}$ \cite{Taschin2013}. However, these experiments were carried out at temperatures above 247~K, which is much higher than the lowest temperatures reported here. By probing the water droplets of the present study in the low-frequency region it would thus be possible to elucidate the nature of the structural evolution and ice formation occurring in liquid water in the deeply supercooled regime.

This work was in part supported by the BMBF through Grant No. 05K13RF5.

\newpage
%--------------------- FIGURES -----------------------\%
%
%\%\%\% Figure 1 \%\%\%\%\%
%
%
%
%\%\%\% Figure 2 \%\%\%\%\%
%
%
%
%\%\%\% Figure 3 \%\%\%\%\%
%
%
%
%\%\%\% Figure 3 \%\%\%\%\%
%


\begin{thebibliography}{00}%{99}

\bibitem{Fahrenheit1724}
D. G. Fahrenheit, Phil. Trans. \textbf{33}, 381 (1724).

\bibitem{Poole1992}
P. H. Poole, F. G. Sciortino, U. Essmann, and H. E. Stanley, Nature \textbf{360}, 324 (1992).

\bibitem{Palmer2014}
J. C. Palmer, F. Martelli, Y. Liu, R. Car, A. Z. Panagiotopoulos, and P. G. Debenedetti, Nature \textbf{510}, 385 (2014).

\bibitem{Moore2011}
E. Moore and V. Molinero, Nature \textbf{479}, 506 (2011).

\bibitem{Russo2014}
J. Russo, F. Romano, and H. Tanaka, Nat. Mater. \textbf{13}, 733 (2014).

\bibitem{Haji-Akbari}
A. Haji-Akbari and P. G. Debenedetti, Proc. Natl. Acad. Sci. \textbf{112}, 10582 (2015).

\bibitem{Rosenfeld2000}
D. Rosenfeld and W. L. Woodley, Nature \textbf{405}, 440 (2000).

\bibitem{Murray2012}
B. J. Murray, D. O'Sullivan, J. D. Atkinson, and M. E. Webb, Chem. Soc. Rev., \textbf{41}, 6519 (2012).

\bibitem{Nilsson2015}
A. Nilsson and G. M. Pettersson, Nat. Commun. \textbf{6}, 8998 (2015).

\bibitem{Debenedetti2003}
P. G. Debenedetti, J. Phys.: Condens. Matter \textbf{15}, R1669 (2003).

\bibitem{Speedy76}
R. J. Speedy and C. A. Angell, J. Chem. Phys. \textbf{65}, 851 (1976).

\bibitem{Dehaoui2015}
A. Dehaoui, B. Issenmann, and F. Caupin, Proc. Natl. Acad. Sci. \textbf{112},12020 (2015).

\bibitem{Gallo2016}
P. Gallo, K. Amann-Winkel, C. A. Angell, M. A. Anisimov, F. Caupin, C. Chakravarty, E. Lascaris, T. Loerting, A. Z. Panagiotopoulos, J. Russo, J. A. Sellberg, H. E. Stanley, H. Tanaka, C. Vega, L. Xu, and L. G. M. Pettersson, Chem. Rev., \textbf{116}, 7463 (2016).

\bibitem{Hare1986}
D. E. Hare and C. M. Sorensen, J. Chem. Phys. \textbf{84}, 5085 (1986).

\bibitem{Taborek1985}
P. Taborek, Phys. Rev. B \textbf{32}, 5902 (1985).

\bibitem{Riechers2013}
B. Riechers, F. Wittbracht, A. H\"utten, and T. Koop, Phys. Chem. Chem. Phys. \textbf{15}, 5873 (2013).

\bibitem{Mallamace2013}
F. Mallamace, C. Corsaro, and H. E. Stanley, Proc. Natl. Acad. Sci. \textbf{110}, 4899 (2013).

\bibitem{Murata2012}
K.-I. Murata and H. Tanaka, Nat. Mater. \textbf{11}, 436 (2012).

\bibitem{Manka2012}
A. Manka, H. Pathak, S. Tanimura, J. W\"olk, R. Strey, and B. E. Wyslouzil, Phys. Chem. Chem. Phys. \textbf{12}, 4505 (2012).

\bibitem{Caupin2015}
F. Caupin, J. Non-Cryst. Solids \textbf{407}, 441 (2015).

\bibitem{Faubel1988}
M. Faubel, S. Schlemmer, and J. P. Toennies, Z. Phys. D \textbf{10}, 269 (1988).

\bibitem{Sellberg2014}
J. A. Sellberg, C. Huang, T. A. McQueen, N. D. Loh, H. Laksmono, D. Schlesinger, R. G. Sierra, D. Nordlund, C. Y. Hampton, D. Starodub, D. P. DePonte, M. Beye, C. Chen, A. V. Martin, A. Barty, K. T. Wikfeldt, T. M. Weiss, C. Caronna, J. Feldkamp, L. B. Skinner, M. M. Seibert, M. Messerschmidt, G. J. Williams, S. Boutet, L. G. M. Pettersson, M. J. Bogan, and A. Nilsson, Nature \textbf{510}, 381 (2014).

\bibitem{Sellberg2015}
J. A. Sellberg, T. A. McQueen, H. Laksmono, S. Schreck, M. Beye, D. P. DePonte, B. Kennedy, D. Nordlund, R. G. Sierra, D. Schlesinger, T. Tokushima, I. Zhovtobriukh, S. Eckert, V. H. Segtnan, H. Ogasawara, K. Kubicek, S. Techert, U. Bergmann, G. L. Dakovski, W. F. Schlotter, Y. Harada, M. J. Bogan, P. Wernet, A. F\"ohlisch, L. G. M. Pettersson, and A. Nilsson, J. Chem. Phys. \textbf{142}, 044505 (2015).

\bibitem{Laksmono2015}
H. Laksmono, T.A. McQueen, J. A. Sellberg, N. D. Loh, C. Huang, D. Schlesinger, R. G. Sierra, C. Y. Hampton, D. Nordlund, M. Beye, A. V. Martin, A. Barty, M. M. Seibert, M. Messerschmidt, G. J. Williams, S. Boutet, K. Amann-Winkel, T. Loerting, L. G. M. Pettersson, M. J. Bogan, and A. Nilsson, J. Phys. Chem. Lett. \textbf{6}, 2826 (2015).

\bibitem{Smith2006}
J. D. Smith, C. D. Cappa, W. S. Drisdell, R. C. Cohen, and R. J. Saykally, J. Am. Chem. Soc. \textbf{128}, 12892 (2006).

\bibitem{Azouzi2013}
M. E. Mekki-Azouzi, C. Ramboz, J.-F. Lenain, and F. Caupin, Nat. Phys. \textbf{9}, 38 (2013).

\bibitem{DePonte2008}
D. P. DePonte, U. Weierstall, K. Schmidt, J. Warner, D. Starodub, J. C. H. Spence, and R. B. Doak, J. Phys. D \textbf{41}, 195505 (2008).

\bibitem{Frohn2000}
A. Frohn and N. Roth, ``Dynamics of Droplets'' (Springer, Berlin, 2000).

\bibitem{Kuehnel2011}
M. K\"uhnel, J. M. Fern\'andez, G. Tejeda, A. Kalinin, S. Montero, and R. E. Grisenti. Phys. Rev. Lett. \textbf{106}, 245301 (2011).

\bibitem{Sun2013}
C. Q. Sun, X. Zhang, X. Fu, W. Zheng, J.-l. Kuo, Y. Zhou, Z. Shen, and J. Zhou, J. Phys. Chem. Lett. \textbf{4}, 3238 (2013).

\bibitem{Thurn1985}
R. Thurn and W. Kiefer, App. Opt. \textbf{24}, 1515 (1985).

\bibitem{Kiefer1996}
W. Kiefer, J. Popp, M. Lankers, M. Trunk, I. Hartmann, E. Urlaub, and J. Musick, J. Mol. Struct. \textbf{408}, 113 (1996). 

\bibitem{Chylek76}
P. Ch\'ylek, J. Opt. Soc. Am. \textbf{66}, 285 (1976).

\bibitem{Chylek78}
P. Ch\'ylek, J. T. Kiehl, and M. K. W. Ko, Phys. Rev. A \textbf{18}, 2229 (1978).

\bibitem{Bohren98}
C. F. Bohren and D. R. Huffman, Absorption and Scattering of Light by Small Particles (Wiley, Berlin, 1998).

\bibitem{Suzuki2012}
H. Suzuki, Y. Matsuzaki, A. Muraoka, and M. Tachikawa, J. Chem. Phys. \textbf{136}, 234508 (2012).

\bibitem{Sun_b2013}
Q. Sun, Chem. Phys. Lett. \textbf{568}, 90 (2013).

\bibitem{Harvey1998}
A. H. Harvey, J. S. Gallagher, and J. M. H. Levelt Sengers, J. Phys. Chem. Ref. Data \textbf{27}, 761 (1998).

\bibitem{Carroll2002}
L. Carroll and M. Henry, Appl. Opt. \textbf{41}, 1330 (2002).

\bibitem{Hare1987}
D. E. Hare and C. M. Sorensen, J. Chem. Phys. \textbf{87}, 4840 (1987).

\bibitem{Murphy2005}
D. M. Murphy and T. Koop, Q. J. R. Meteorol. Soc. \textbf{131},1539 (2005).

\bibitem{Kalova2010}
J. Kalova and R. Mares, Int. J. Themophys. \textbf{31}, 756 (2010).

\bibitem{Walrafen1967}
G. E. Walrafen, J. Chem. Phys. \textbf{47}, 114 (1967).

\bibitem{Aliotta2014}
F. Aliotta, P. V. Giaquinta, R. C. Ponterio, S. Prestipino, F. Saija, G. Salvato, and C. Vasi, Sci. Rep. \textbf{4}, 7230 (2014).

\bibitem{Pathak2016}
H. Pathak, J. C. Palmer, D. Schlesinger, K. T. Wikfeldt, J. A. Sellberg, L. G. M. Pettersson, and A. Nilsson, J. Chem. Phys. \textbf{145}, 134507 (2016).

\bibitem{Bakker2010}
H. J. Bakker and J. L. Skinner, Chem. Rev. \textbf{110}, 1498 (2010).

\bibitem{Hare1990}
D. E. Hare and C. M. Sorensen, J. Chem. Phys. \textbf{93}, 25 (1990).

\bibitem{Taschin2013}
A. Taschin, P. Bartolini, R. Eramo, R. Righini, and R. Torre, Nat. Commun. \textbf{4}, 2401 (2013).

\end{thebibliography}
\end{document}